\documentclass[aps,prx,onecolumn,superscriptaddress,nofootinbib,10pt]{revtex4-2}\usepackage{graphics}
\usepackage{amsmath}
\usepackage{physics}
\usepackage{graphicx}
\usepackage{color}
\usepackage{amsfonts,wasysym}
\usepackage{amssymb}
\usepackage{tikz}
\usepackage{placeins}
\usepackage{comment}
\usepackage{acronym}
\usepackage{hyperref}

\FloatBarrier

\newcommand{\bs}[1]{\boldsymbol{#1}}

\definecolor{MK}{rgb}{0.0,0.0,0.9}

\definecolor{EX}{rgb}{0.5,0.5,0.0}

\definecolor{SuggestionHighlight}{rgb}{0.5,0.0,0.5}

\definecolor{ACcomm}{rgb}{0.0,0.9,0.1}
\definecolor{LB}{RGB}{134,41,198}%fiorentina color

\DeclareMathOperator*{\argmax}{arg\,max}

\begin{document}

% acronyms 
\newacro{MPS}{Matrix Product States}
\newacro{MPO}{Matrix Product Operators}
\newacro{DMRG}{Density Matrix Renormalization Group}
\newacro{SPT}{Symmetry-Protected Topological }
\newacro{SVM}{Support Vector Machine}
\newacro{QCNN}{Quantum Convolutional Neural Network}

\title{Learning Topological Quantum Phases from Limited Subsystems} 

\author{Mehran Khosrojerdi}\email{mehran.khosrojerdi@unifi.it}
\affiliation{Department of Physics and Astronomy, University of Florence,
via G. Sansone 1, I-50019 Sesto Fiorentino (FI), Italy}
 
\author{Sougato Bose}
\affiliation{Department of Physics and Astronomy, University College London,
 Gower Street, London WC1E 6BT, United Kingdom}

\author{Alessandro Cuccoli} \affiliation{Department of Physics and 
Astronomy, University of Florence, via G. Sansone 1, I-50019 Sesto 
Fiorentino (FI), Italy} \affiliation{ INFN Sezione di Firenze, via G. 
Sansone 1, I-50019, Sesto Fiorentino (FI), Italy } 

\author{Paola Verrucchi} 
\affiliation{ISC-CNR, UOS Dipartimento di Fisica, Universit\`a di Firenze, I-50019, Sesto Fiorentino (FI), Italy}
\affiliation{Department of Physics and Astronomy, University of Florence, via
G. Sansone 1, I-50019 Sesto Fiorentino (FI), Italy} 
\affiliation{ INFN Sezione di Firenze, via G. Sansone 1, I-50019, Sesto 
Fiorentino (FI), Italy } 

\author{Abolfazl Bayat}
\affiliation{Institute of Fundamental and Frontier Sciences, University of Electronic Science and Technology of China, Chengdu 611731, China}

\author{Leonardo Banchi}
\affiliation{Department of Physics and Astronomy, University of Florence,
via G. Sansone 1, I-50019 Sesto Fiorentino (FI), Italy}
\affiliation{ INFN Sezione di Firenze, via G. Sansone 1, I-50019, Sesto Fiorentino (FI), Italy }

\date{\today}
\begin{abstract}
Characterizing quantum topological phases requires measuring non-local string order parameters, demanding access to the full system, which is  often experimentally unfeasible. In this work, we introduce a data-efficient supervised learning framework that circumvents this limitation by recognizing quantum phases from small subsystems. Our protocol utilizes a quantum kernel constructed from the reduced density matrices of these subsystems, which can be efficiently estimated experimentally. We benchmark our framework  with the classification of  the phase diagrams of two spin models on one-dimensional lattices, namely the generalized cluster-Ising spin-1/2 chain and the anisotropic Haldane spin-1 chain. Remarkably, our approach achieves high accuracy in phase classification when operations are limited to as few as one to four sites, and it also generalizes to longer chains even when trained on moderate system sizes. 
These findings demonstrate that local reduced density matrices preserve vital signatures of global topological phases, offering a practical route to characterize rich phase diagrams of quantum many-body systems. 
\end{abstract}

\maketitle

\section{Introduction} 
\label{s.introduction}

Drawing quantum phase diagrams of one-dimensional spin systems relies
on well-established numerical and analytical techniques~
\cite{sachdev1999quantum,vojta2003quantum,mendes2019entanglement,calabrese2008entanglement,calabrese2004entanglement,calabrese2009entanglement,bayat2010negativity,santos2011negativity,calabrese2012entanglement,ruggiero2016negativity,chen2003ground,langari2013ground,ejima2015comparative,yu2024quantum,koyluouglu2024measuring,albuquerque2009quantum}. 
Continuous second order phase transitions are described by the Landau-Ginzburg paradigm, where
 symmetry breaking marks the phase transition and different phases are characterized in terms of local 
 order parameters.
However, the Landau-Ginzburg paradigm fails to identify phases of matter 
which are accompanied by topological properties. The key examples are
\ac{SPT} phases \cite{smacchia2011statistical,son2011quantum}, in which  symmetry breaking is absent and no local order parameter can 
be used to identify them. 
In fact, in order to identify an \ac{SPT} phase one requires the measurement of a
non-local observable whose expectation value is the string order parameter. 
 Such measurement is difficult in practice as it demands access to the entire system.
Motivated by such difficulties, many methods have been invented to efficiently characterize topological phases
\cite{perez2008string,pollmann2010entanglement,ohta2016topological, 
ohta2015phase,smacchia2011statistical, son2011quantum, 
PhysRevB.96.060404, tzeng2008scaling,deng2017machineLearning,Lian2019machinelearning,zhang2022experimental,zhang2022digital,xiang2024long}.

Recently, machine-learning-based methods have gained traction for classifying the phases of quantum many-body systems.
In a broad picture, these methods can be divided into two main categories. 
In the first category, the task is achieved by feeding the outcomes from fixed measurements directly into a classical neural network~\cite{carrasquilla2017machine, 
Cristianini_Shawe-Taylor_2000, schuld2021supervised,huang2022provably, 
khosrojerdi2025unsupervised, wu2024learning,du2025artificial,wu2024learning,wu2024Efficient}.
In the second category, quantum machine learning is employed using variational quantum algorithms
\cite{cong2019quantum,li2024ensemble,lazzarin2022multi,gong2023quantum,mansuroglu2023variational,tepaske2023optimal,mc2023classically}, 
principal component analysis \cite{lloyd2014quantum,banchi2025statistical}, 
\acp{SVM}~\cite{Khosrojerdi_2025, parigi2025supervised} and other kernel methods \cite{sancho2022quantum,wu2023quantum}.
Despite their success, all these methods come with their own challenges. For
instance, variational quantum algorithms 
have trainability issues for large system sizes due to barren plateau
\cite{mcclean2018barren}, while principal component analysis requires fault tolerant quantum
hardware

The estimation of string order parameters and previous applications of quantum machine learning methods demand access 
to the entire system. This raises two main questions: i) while full characterization of topological phases requires global information, 
is it really necessary to have global access just for phase identification? ii) can we exploit quantum machine learning methods with locally extracted information to this aim? 
%Can we identify topological phases just by partial access to the system? At first sight, this might sound impossible as, by definition, there is no local order parameter. 
In this paper, we address these questions by developing a method which only demands access to a small 
number of sites in a many-body system. Central to our approach is the use of kernels obtained from reduced density matrices, which is then 
fed into a classical \ac{SVM} classifier. Such kernels can be experimentally estimated directly in hardware through different techniques, including the swap test \cite{schuld2021supervised}, single-triplet measurements 
\cite{banchi2016entanglement}, shadow \cite{huang2020predicting} and ultimately full tomography \cite{haah2016sample}. 
We apply our method to two models that display rich phase diagrams, each including an \ac{SPT} phase, namely the generalized cluster-Ising spin-$\frac12$ chain and the anisotropic Haldane spin-1 chain. 
In both models, our protocol can identify different phases of the diagram with high accuracy with access to just 4 sites. Interestingly, we find that, to achieve such accuracies, what matters is the size of the accessible section of the chain, not the size of the whole chain. Indeed, models trained with chains of moderate sizes can directly generalize to larger system sizes. 
Our results show that phase 
recognition can indeed be achieved referring solely to subsystems, which
highlights the practical feasibility and the conceptual 
significance of the proposed approach.

%structure of specific quantum states, namely the ground states to be 
%assigned to the different quantum phases, states that belong to Hilbert 
%spaces with huge dimension, even for relatively short
%one-dimensional spin systems,  or spin-{\it chains} for short. This 
%poses a significant challenge not only to numerical 
%approaches but also to experimental ones, where it requires a demanding 
%quantum tomography.
%Aim of this work is that of providing a method for making the study of 
%quantum phase diagrams in many-body systems more accessible.
%We propose to perform quantum phase recognition based 
%on the properties of subsystems that, despite being significantly 
%smaller than the original system, retain signatures of the global 
%phases. We focus on phase recognition in gapped regimes, leaving the 
%characterization of gapless critical points outside the scope of 
%this work, and apply the method to the Cluster Ising model, which features a
%\ac{SPT} phases, and to the Haldane anisotropic spin-1 chain, a 
%canonical model for topological order. 

The article is structured as follows. 
Section~\ref{s.method} presents the methodology, detailing the 
construction of the dataset and the machine learning framework.
Section~\ref{s.results} reports and analyzes the results obtained from 
our 
approach. Finally, Section~\ref{s.conclusion} summarizes the findings 
and 
highlights the main achievements and insights of this work.

\section{Support Vector Machines with Quantum Kernels}%
\label{s.method}

In this paper, we focus on phase discrimination in many-body systems prepared in their ground state, assuming that our operational access is restricted to only a few sites. While this limited access may suffice for recognizing phases characterized by a local order parameter, it poses significant challenges for identifying topological phases, whose order parameter is constructed from nonlocal string operators. 
Since quantum phases of matter are only well-defined in large systems, ideally in the thermodynamic limit, we use \ac{MPS} representations~\cite{schollwock2011density} of the ground states to simulate long chains that lie well beyond the reach of exact diagonalization. Specifically, we make use of the Python libraries reported in Refs.~\cite{gray2018quimb}. 

%Our supervised-learning method uses a dataset of classical representations of different ground states, each corresponding to specific values of the parameters entering the Hamiltonian of the system we aim at studying. Getting representations of these ground states is a hard task, due to the exponential growth of the Hilbert-space dimension with the system size. However, numerical libraries \cite{gray2018quimb,fishman2022itensor} exist, based on tensor network methods (specifically \acp{MPS} formalism and  \ac{DMRG} algorithms \cite{schollwock2011density}) that allow one to get accurate knowledge of the ground states. As we deal with quantum-phase recognition, here ``accurate'' means either complete \cite{Khosrojerdi_2025} or anyhow detailed enough to provide quantities that distinguish between different phases of matter. These may include order parameters \cite{carrasquilla2017machine,PhysRevResearch.5.013082}, correlation functions \cite{PhysRevB.107.214451}, entanglement entropy \cite{PhysRevB.96.060404, physRevLett.96.110405,kitaev2006topological}, and others.

\begin{figure}[t]
  \begin{center}
    \centering
    \begin{minipage}[c]{0.54\textwidth}
      \centering

      % Panel (a)
      \includegraphics[width=0.8\linewidth]{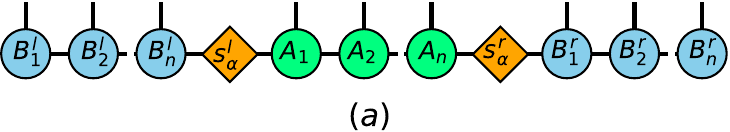}

      \vspace{0.5em}

      % Panel (b)
      \includegraphics[width=0.8\linewidth]{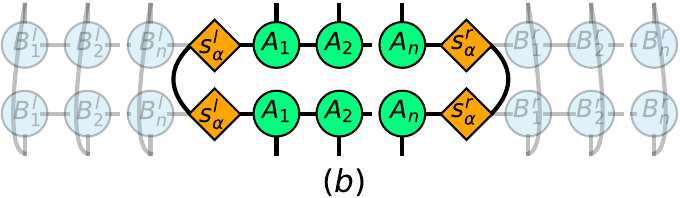}
    \end{minipage}
    \hfill
    \begin{minipage}[c]{0.42\textwidth}
      \centering

      % Panel (c)
      \includegraphics[width=0.8\linewidth]{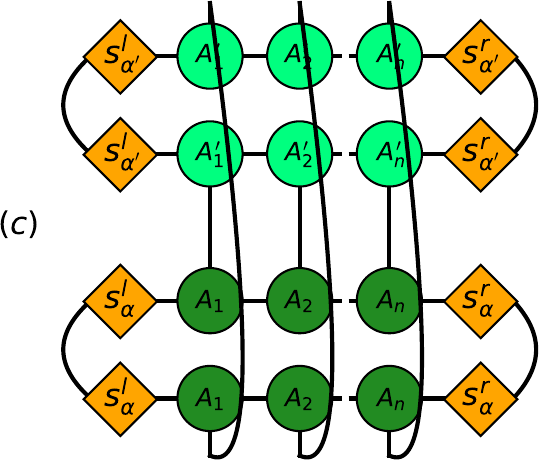}
    \end{minipage}
    \caption{(a) Bipartite \ac{MPS} in canonical form. The chain is split into a block of adjacent sites $A$ and a block $B$ containing the rest of the system, with the bond index $\alpha$ representing the Schmidt decomposition Eq.~\eqref{eq:schmidt} and encoding the entanglement across the cut.
    (b) Reduced density matrix $\rho_A$ of block $A$, obtained by tracing out over block $B$ in the \ac{MPS} Schmidt decomposition. (c) Kernel \eqref{eq:kernel} between two reduced density matrices, measuring their similarity. }
    \label{fig:canonical_MPS}
  \end{center}
\end{figure}

Let's consider quantum many-body systems on a lattice whose 
Hamiltonians $\cal{H}$ depend on a certain number of parameters, 
$(x_1,x_2...)\equiv\bs{x}$, e.g. external fields, anisotropies, and couplings.
By varying $\bs{x}$ 
within a chosen range, we construct a set $\{\mathcal{H}(\bs{x})\}$ of 
Hamiltonians, each corresponding to a specific point in the 
parameter space and hence to a distinct phase. The ground state $\ket{\psi(\bs x)}$ of each Hamiltonian $\mathcal{H}(\bs{x})$  is obtained through 
an \ac{MPS} simulation. Since we assume that our access to the system is limited to a few sites, we split the system into two parts, say blocks $A$ and $B$. We can perform our operations only on sites within the block $A$ while the rest of the system, i.e. block $B$, is not accessible. 
In the most general setting shown in Fig.~\ref{fig:canonical_MPS} the block $A$ can 
be in the middle of the chain and a generic state has a tripartite structure $B_l\cup A\cup B_r$ 
where the left and right parts make up the block $B=B_l\cup B_r$. Therefore, a general (ground) state 
can be expanded into orthonormal bases as 
\begin{equation}
  \ket{\psi(\bs{x}_n)} = \sum_{\alpha\beta_l\beta_r} \Psi_{\beta_l\alpha\beta_r} \ket{\beta_l}_{B_l} \ket{\alpha}_A \ket{\beta_r}_{B_r}.
    \label{eq:schmidt}
\end{equation}
For such states we can still perform two Schmidt decompositions. If we first group the indices $\alpha$ and $\beta_r$ together and perform 
a singular value transformation on the resulting matrix we 
get $\Psi_{\beta_l\alpha\beta_r} = \sum_{\gamma} U_{\beta_l\gamma} s^l_\gamma V_{\gamma,\alpha\beta_r}$, where $s_\gamma^l$ are 
the Schmidt coefficients, ordered in decreasing order,  and $U$ and $V$ are unitary matrix. In simulations, the extra index $\gamma$ is 
truncated to a fixed maximum bond dimension $\chi$.  The same approach can now be performed from the right, grouping the indices 
$\gamma$ and $\alpha$ together to write $V_{\gamma\alpha,\beta_r} = \sum_{\delta}  \Gamma_{\gamma\alpha,\delta} s_\delta^r U'_{\delta,\beta_r}$
where $s_\alpha^r$ are other Schmidt coefficients and $U'$ is unitary. Therefore, the Schmidt coefficients $s_{\alpha}^{l/r}$ respectively 
captures the entanglement between $B_l$ and $A\cup B_r$, and between $B_l\cup A$ and $B_r$. 
The same procedure can be iteratively applied spin by spin, starting from the left and from the right, to get the so called ``mixed canonical form'' \cite{schollwock2011density} shown in  Fig.~\ref{fig:canonical_MPS}(a) where, crucially, the tensors $B_i^l$ and $B_i^r$ belong to unitary matrices, 
and each index corresponding to a horizontal line in Fig.~\ref{fig:canonical_MPS} is truncated to $\chi$. 

Since the tensors in the left and right blocks are unitary, when tracing over $B$ to get the reduced 
density matrix $\rho_A(\bs x_n)$ they disappear, and only the contraction over 
the Schmidt vectors remain, as shown in Fig.~\ref{fig:canonical_MPS}(b). 
%An appropriate 
%choice of the bipartition should guarantee that enough 
%information is retained in the smallest one, say $A$, for identifying 
%different quantum phases.
%The bipartite \ac{MPS} in canonical form provides the decomposition
%\begin{equation}
%    \ket{\psi(\bs{x}_n)} = \sum_{\alpha = 1}^{\chi} s_\alpha 
%\ket{A_\alpha}\otimes\ket{B_\alpha}~,
%\end{equation}
%where $\{\ket{A_\alpha}\}$ and $\{\ket{B_\alpha}\}$ are orthonormal states 
%for $A$ and $B$, respectively, and $\chi$ is the bipartite MPS bond 
%dimension at the cut between sites in block A and B (see figure 
%\ref{fig:canonical_MPS}~(a)); this parameter is fixed to some suitable value which is much smaller than 
%the Schmidt rank, as the latter can grow exponentially. 
%%with the equality meaning that Eq.~\eqref{eq:schmidt} is the Schmidt 
%%decomposition of the DMRG ground state.
%The state of $A$ reads
%\begin{equation}
%	\rho_A(\bs{x}_n) = 
%\mathrm{Tr}_B\big[\ket{\psi(\bs{x}_n)}\!\!\bra{\psi(\bs{x}_n)}\big] = 
%\sum_{\alpha=1}^{\chi} s_\alpha^2 \ket{A_\alpha}\!\!\bra{A_\alpha}
%~, 
%\end{equation}
Given two states $\rho_A(\bs{x}_n)$ and $\rho_A(\bs{x}_m)$ obtained with the above procedure, we 
can now quantify their similarity by the inner product
\begin{align}
\label{eq:kernel}
  K(\bs{x}_n, &\bs{x}_m) = \mathrm{Tr} \big[ \rho_A(\bs{x}_n) \, 
\rho_A(\bs{x}_m) \big], % = \\ &= 
%\sum_{\alpha, \alpha'} 
%s_\alpha^2(\bs{x}_n) \, s_{\alpha'}^2(\bs{x}_m) \,
%   \left| \langle A_\alpha(\bs{x}_n) | A_{\alpha'}(\bs{x}_m) \rangle 
%   \right|^2, %\nonumber
\end{align}
that defines the kernel $K_{nm}$ used as input of the classical learning 
algorithm. The tensor contractions to construct those kernel elements are shown in  
Figure~\ref{fig:canonical_MPS}\,(c). 
Notice that $\rho_A$ is not, in general, a projector and the diagonal elements of the kernel, $K(\bs{x}_n, \bs{x}_n) = 
\mathrm{Tr}[\rho_A(\bs{x}_n)^2]$, are consequently less than one.

Once the kernel matrix is evaluated, we use it as input for the 
\ac{SVM} algorithm~\cite{Cristianini_Shawe-Taylor_2000}, a well-established supervised 
classification method that aims at finding an
optimal hyperplane in the ``feature'' space. For a feature space of density matrices
and  a ``decision function'' 
\begin{equation}
    f(\bs{x}) = \mathrm{Tr}[W \rho_A(\bs{x})] + b,
    \label{eq:decision}
\end{equation}
where $W$ is an observable and $b$ is a shift, the hyperplane is defined as $f(\bs x)=0$. 
A single \ac{SVM} solves a binary classification problem,
where the predicted class is given by the sign of $f(\bs{x})$; 
multi-class problems are split into
multiple binary classifications. 

Training consists of optimizing over $W$ and $b$ given a training set with $M$ pairs $(\bs x_i,y_i)$, where $y_i$ is the 
true label of $\bs x_i$. 
Since $\rho_A(\bs x)$ may live in an exponentially large Hilbert space, the optimization over $W$ may be intractable. 
On the other hand, using the representer theorem from \ac{SVM} \cite{schuld2021supervised},
$W$ can be expressed solely in terms of $M$ coefficients $\alpha_i$ such that
$W= \sum_{i=1}^M \alpha_i y_i \rho(\bs x_i)$. 
It can be shown \cite{Cristianini_Shawe-Taylor_2000}
that the resulting optimization problem for the coefficients $\alpha_i$ and $b$ is convex and results in 
\begin{align}
    \argmax_{0 \leq \alpha_i < C} \left[ \sum_{i=1}^M \alpha_i - 
    \frac{1}{2} \sum_{i,j=1}^M y_i y_j \alpha_i \alpha_j K(\bs{x}_i, \bs{x}_j) \right],
    \label{eq:svm opt}
\end{align}
with linear constraint 
$\sum_{i=1}^M \alpha_i y_i = 0$, where $C$ is a regularization hyperparameter. In our simulations $C$ is kept to its default value $C=1$.  
Plugging $W$ in the decision function Eq.~\eqref{eq:decision} and using Eq.~\eqref{eq:kernel}, 
the predicted value of the label $y$ for a new test point $\bs x$  is 
\begin{align}
  y &= \mathrm{sign}[f(\bs x)]= 
    \mathrm{sign} \left[ \sum_{i=1}^M \alpha_i y_i K(\bs{x}_i, \bs{x}) + b \right],
    &
\mathrm{where~~~}  b &=  \sum_{i=1}^M \frac{y_i - \sum_{j=1}^M \alpha_j y_j K(\bs x_i,\bs x_j)}{M}.
    \label{eq:y predicted}
\end{align}
Note that the labeling only depends on the kernel between $\bs x$ and all training points $\bs x_i$.

As already mentioned, the kernel of Eq.~\eqref{eq:kernel} can be experimentally estimated in different ways, 
including the swap test \cite{schuld2021supervised}, single-triplet measurements 
\cite{banchi2016entanglement}, shadow \cite{huang2020predicting} and full tomography \cite{haah2016sample}. 
However, having obtained the coefficients $\alpha_i$ through the optimization in \eqref{eq:svm opt},
if block $A$ contains few spins  then $W$ can also be expressed in the spin basis as 
\begin{equation}
  W = \sum_{i=1}^M \alpha_i y_i \rho(\bs x_i) = \sum_{k=1}^{d_A^2} c_k P_k.
\end{equation}
where $d_A$ is the dimension of the Hilbert space of $\rho_A$ and 
$P_k$ are spin operators. For example in the case of qubits $d_A=2^{N_A}$ whenre $N_A$ is the number of spins 
in block $A$ and $P_k$ are products of Pauli operators. 
The coefficients $c_k$ can be obtained from $\alpha_i$ via $c_k = \Tr[ W P_k] / d_A$. 
When $d_A$ is moderately small, it would be more efficient to compute the label of a new test point $\bs x$ by directly measuring
$\Tr[\rho(\bs x) P_k]$ and then estimate the label $y$ through the sign of $f(\bs x)$ in Eq.~\eqref{eq:decision}.

\section{Results\label{s.results}}

We have benchmarked our ML quantum-phase classification method against 
two quantum spin chains, namely the spin-$1/2$ Generalized 
Cluster-Ising (GCI) model and the spin-$1$ anisotropic Haldane model. The 
$T=0$ properties of both systems 
%extensively enter the ML literature related to phase recognition~\cite{???}, as they 
feature highly non-trivial quantum phase diagrams, with a SPT phase marked by a 
finite non-local string-order parameter that detects the underlying 
topological entanglement. %... \cite{cong2019quantum, SoneTY_2024,Khosrojerdi_2025,khosrojerdi2025unsupervised}. 

\subsection{Generalized Cluster-Ising spin-$\frac12$ chain}

\begin{figure}[t]
	\begin{center}
	    \begin{minipage}[c]{0.4\textwidth}
        \centering
        \includegraphics[ width=0.9\linewidth]{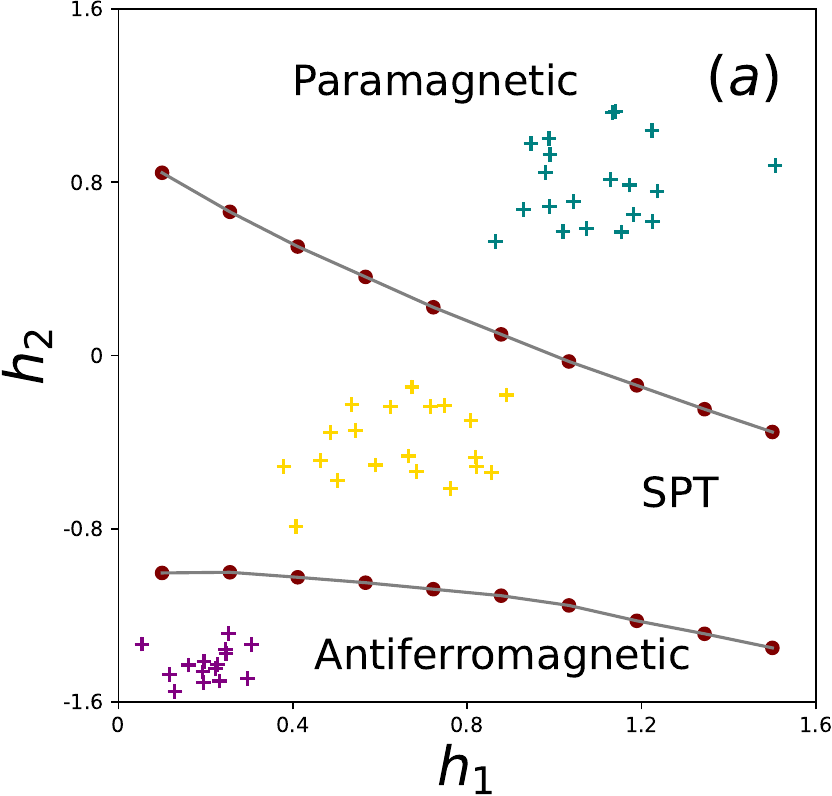}
    \end{minipage}
    %\hfill
    \begin{minipage}[c]{0.4\textwidth}
        \centering
        \includegraphics[ width=0.9\linewidth]{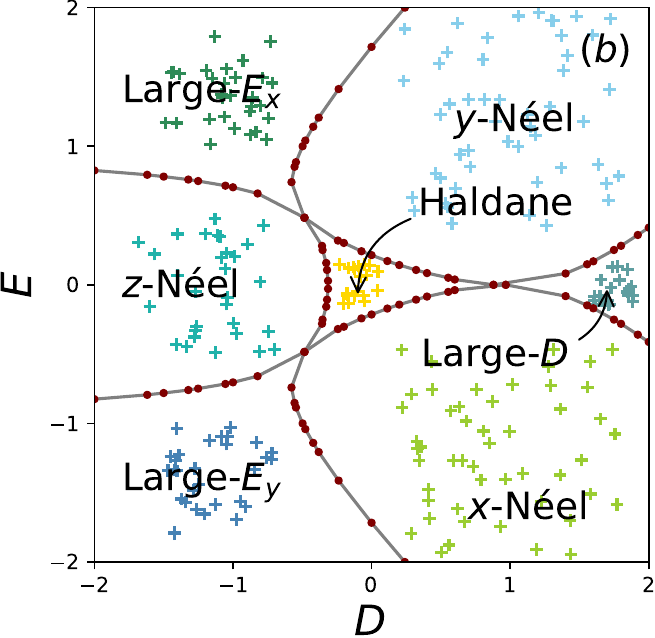}
    \end{minipage}
        \caption{(a) Phase diagram of the Generalized Cluster-Ising model, Eq.~\eqref{e.GCI}: phase-boundary lines are the results of infinite-size DMRG calculations reported in Refs.~\cite{cong2019quantum,SoneTY_2024}. Colored plus signs indicate the training points scattered across the phase diagram, representing the corresponding phase labels. (b) Phase diagram of the Haldane anisotropic spin-$1$ model Eq.~\eqref{eq:Haldane_anisotropic_hamiltonian} as obtained in Ref.\cite{PhysRevB.96.060404}. Colored plus signs indicate the training points scattered across the phase diagram, representing the corresponding phase labels. 
}
	\label{fig:Cluster_Ising_model}
	\end{center}
\end{figure}

The Hamiltonian of the GCI model can be written as
\begin{equation}
    \mathcal{H} = 
    -J\left(\sum_{i=1}^{L-2} Z_{i} X_{i+1} Z_{i+2} + h_1 \sum_{i=1}^L X_i + h_2 \sum_{i=1}^{L-1} X_i X_{i+1}  \right)~,
\label{e.GCI}
\end{equation}
where $i$ runs over the sites of the chain, while
$X_i$ and $Z_i$ are the Pauli operators on site $i$. The 
first term is the so-called Cluster interaction, 
whose strength $J>0$ sets the energy scale, and will be hereafter fixed 
equal to unity; the local term accounts for a magnetic field $h_1$
along the $X$ direction,  while $h_2$ gauges the $XX$ Ising
coupling that can be either ferro-, $h_2>0$, or antiferromagnetic 
$h_2<0$, and the (anti)ferromagnetism is understood along 
the $x$ direction.
The cluster interaction pushes the spins off the $xy$ plane, while the 
other terms favor the in-plane alignment.
The standard symmetry-broken antiferromagnetic phase occurs for $h_2$ 
less than $-1$, all the more so the larger the magnetic field $h_1$, as 
seen in the quantum phase diagram of Fig.~\ref{fig:Cluster_Ising_model}(a).
The SPT phase sets in a region where $h_2\gtrsim -1$, 
consistently with its emergence from the competition between the cluster 
and the antiferromagnetic Ising interaction, and  features
a broken $\mathbb{Z}_2 \otimes \mathbb{Z}_2$ symmetry, 
marked by the onset of a non-local string-order parameter.
Above the SPT phase lies a region with no magnetic correlations, usually 
referred to as paramagnetic phase.
Different approaches, there included infinite-size MPS analysis~\cite{pollmann2012detection}, 
variational quantum algorithms 
~\cite{cong2019quantum,SoneTY_2024,li2024ensemble}, and 
ML approaches~\cite{Khosrojerdi_2025,khosrojerdi2025unsupervised}, 
consistently provide the GCI phase diagram shown in
Fig.~\ref{fig:Cluster_Ising_model}(a).

In this work we introduce the GCI as a benchmark to test the reliability 
of the phase-diagram obtained by the supervised learning approach with 
local observables introduced in Sec.~\ref{s.method}. The training set is depicted via colored plus signs in Fig.~\ref{fig:Cluster_Ising_model}(a). 
These points are randomly chosen across the phase diagram to cover the 
three expected phases, and are labeled accordingly. 
The ground states are obtained by MPS using $\chi=150$.
The \ac{SVM} classification method is implemented using the kernel in Eq.~\eqref{eq:kernel}, with section $A$ made of just a few sites and 
$\rho_A$ obtained from the ground states of the entire chain, via the 
appropriate partial trace.
Testing is performed across the phase diagram using 
Eq.~\eqref{eq:kernel} with the same sections $A$ considered in the 
training process. 
A key advantage of this choice is that, regardless of the 
system size in the training or testing set, the matrices
$\rho_A$ have the same fixed dimension. 
This ensures consistency in kernel construction and allows phase 
classification to be performed efficiently across different system 
sizes. The test data points are taken from a regular grid across the entire phase diagram shown in Fig.~\ref{fig:Cluster_Ising_model}(a). 

\begin{figure*}[t]
	\centering
  \includegraphics[width=1\textwidth]{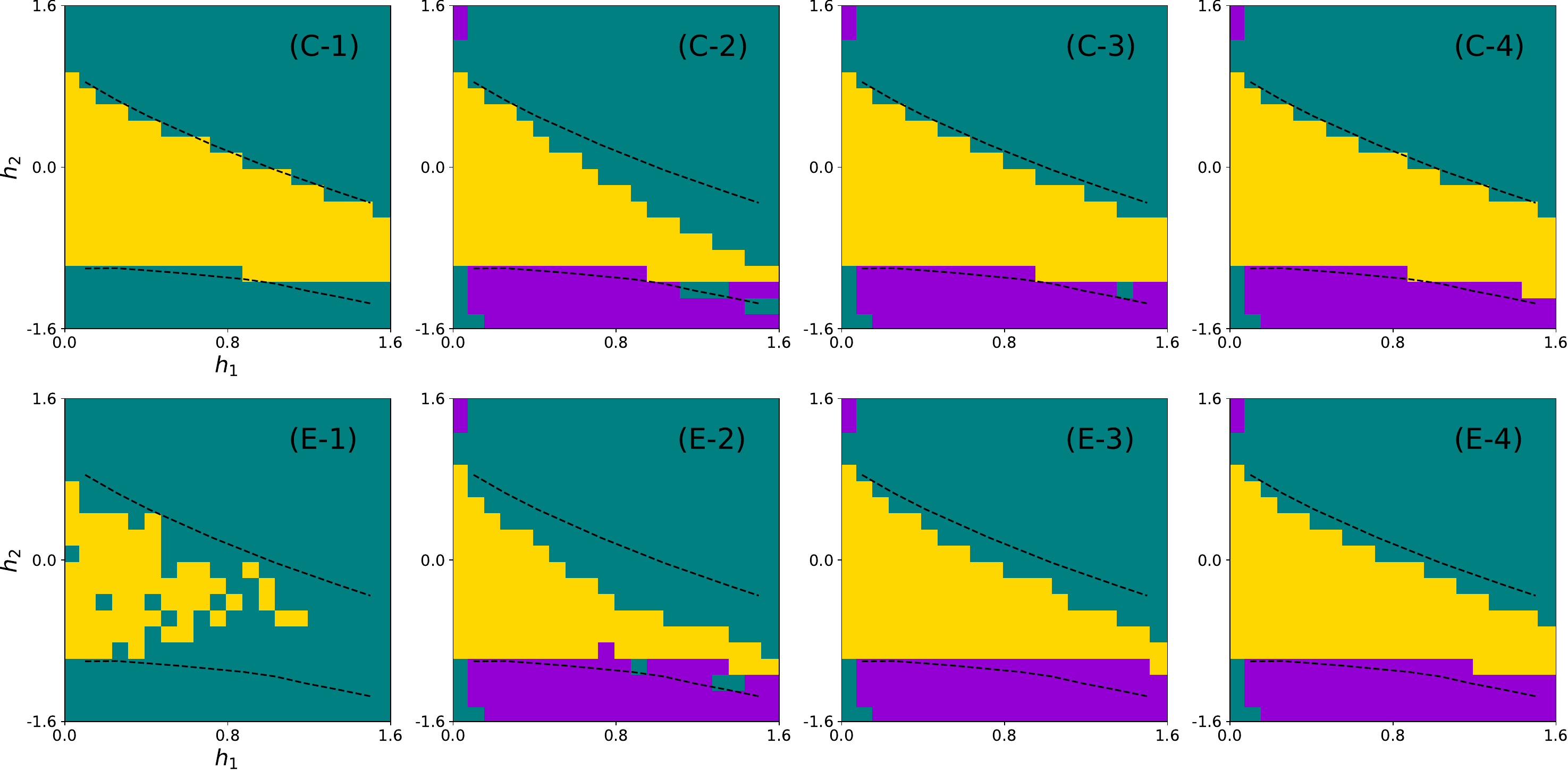}
  \caption{Phase diagram reconstruction of the Generalized Cluster-Ising model based on reduced density matrices for blocks of increasing number of sites. In the upper panels the block is at the center of the chain and (C-$k$) refers to keeping $k$ sites. In the lower panels the block is at the edge of the chain and (E-$k$) refers to keeping $k$ sites. A chain of 51 sites is used both for training and testing. 
 % Reconstruction of the phase diagram of the Generalized Cluster-Ising model shown in Fig.~\ref{fig:Cluster_Ising_model}(a). Effect of keeping an increasing number of sites while tracing out the rest of the system. In the upper panel we keep the sites from the middle of the chain, and (C-$k$) refers to keeping $k$ sites. In the lower panel we keep the sites from an edge of the chain and (E-$k$) refers to keeping $k$ sites. For both training and testing  the full chain consists of 51 sites. 
  }
  \label{fig:Cluster Ising - 51}
\end{figure*}
\begin{figure*}[th!]
	\centering
	\includegraphics[width=1\textwidth]{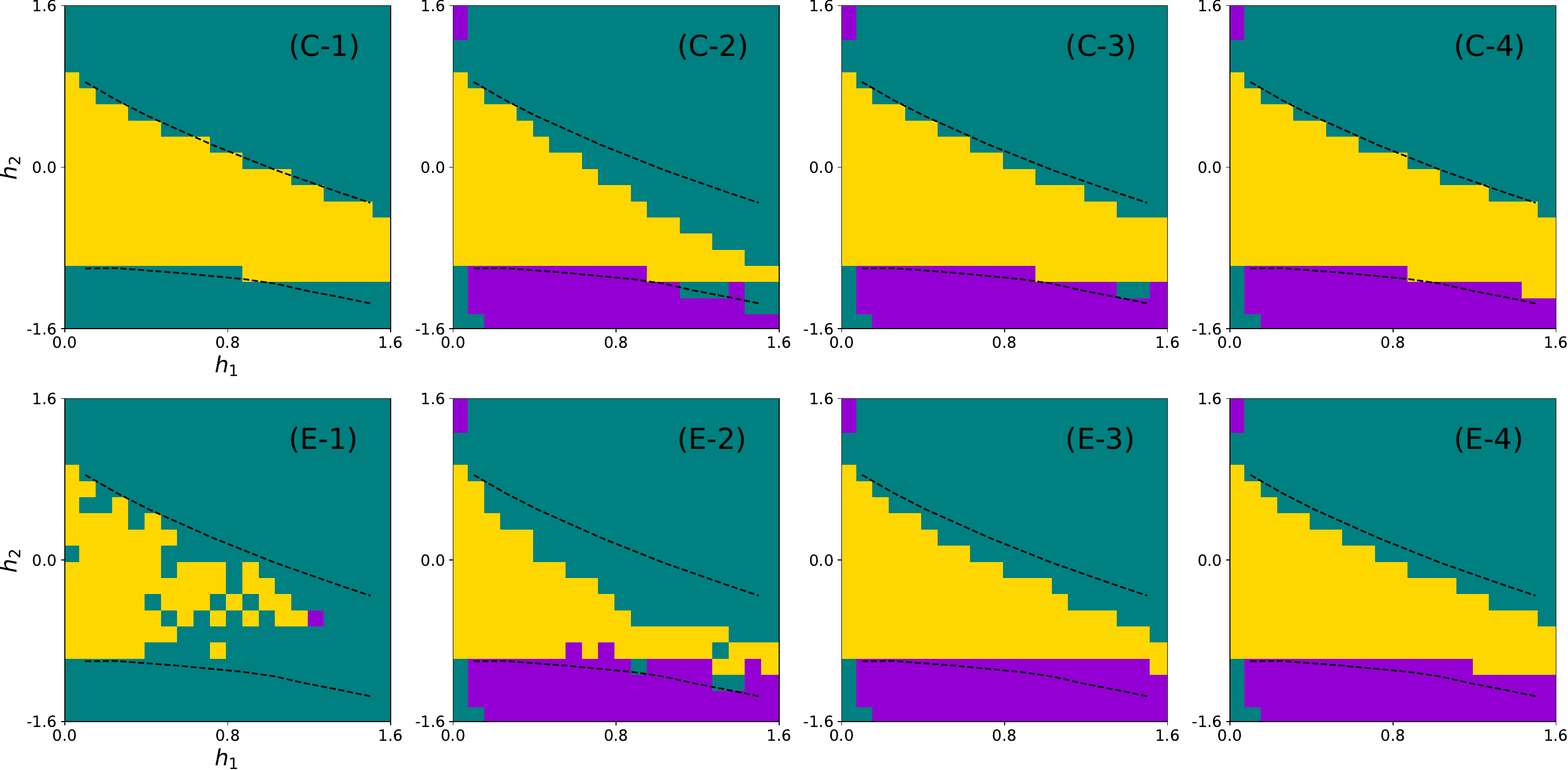}
	\caption{
Reconstruction of the phase diagram of the Generalized Cluster-Ising model as in Fig.~\ref{fig:Cluster Ising - 51}; while a chain of 51 sites is still used for testing, training is performed with a shorter chain of 31 sites. 
  }
	\label{fig:Cluster Ising - 31}
\end{figure*}

Fig.~\ref{fig:Cluster Ising - 51} shows the results obtained for 
the phase diagram of a chain with 51 sites: the classification relies 
on the kernel introduced in Eq.~\eqref{eq:kernel} with sections $A$ 
of 1 to 4 (left to right) sites placed either at the center
(panels (C-1) to (C-4)) or at the edge (panels (E-1) to (E-4)) of the chain.
As expected, when $A$ is at the center, the performance improves as $A$ increases 
in length (see panels (C-1) to (C-4) in Fig.~\ref{fig:Cluster Ising - 51}).
Interestingly, the separation 
between SPT and the other  phases emerges even if just a single
central spin is considered, as seen  in panel (C-1). In this same case, though, the 
antiferromagnetic phase is totally mistaken for the paramagnetic one. 
With $A$ made of two central spins the antiferromagnetic phase shows up, 
but the upper boundary of the SPT phase deviates significantly 
from the expected behavior. With three spins 
the phases are accurately identified, and with four spins they 
are robustly resolved.
When $A$ is at one edge of the chain, see panels (E-1) to (E-4) in Fig.~\ref{fig:Cluster Ising - 51},
the single-spin case is quite useless, 
with most of the phase diagram wrongly labeled. On the other hand, with two spins the phases are already recognized,
and results similar to those obtained with a central block are 
recovered for $3$ and $4$ spins.

We now check if the results of training with smaller system sizes can be used to test longer chains, as  this can reduce the computational cost of the learning process.
To this aim, for the training we take $\rho_A$ from the ground state of a chain with total size $L=31$, while for the test we take $\rho_A$ from  a chain of size  $L=51$.
Fig.~\ref{fig:Cluster Ising - 31} shows the resulting phase 
diagrams in the same way as in Fig.~\ref{fig:Cluster Ising - 51}. The absence of significant differences between Figs.~\ref{fig:Cluster 
Ising - 51} and \ref{fig:Cluster Ising - 31} reveals that the size of 
the chain whose ground states are used to perform the training can 
be safely reduced without significantly affecting the outcome. The reason is that the reduced density matrix $\rho_A$, for any given $N_A$ sites in the block $A$, converges to its thermodynamic equilibrium quite rapidly as the chain size $L$ increases.

%First row of figure \ref{fig:Cluster Ising - 31} exhibits behavior very 
%similar to that of first row of figure \ref{fig:Cluster Ising - 51}. 
%Even with a reduced number of sites in the training set, the overall 
%behavior remains largely consistent.

\subsection{Anisotropic Haldane spin-1 chain}

In this Section we consider the spin-1 model in the family of the  
anisotropic Haldane chains whose Hamiltonian reads 
\cite{PhysRevB.96.060404,ren2018quantum,nico2023entanglement}, 

\begin{equation}
    \mathcal{H}  = J \left( \sum_{i=1}^{L-1} \bs{S}_{i}\cdot\bs{S}_{i+1} 
+ D \sum_{i = 1}^{L}(S^z_i)^2 
+ E \sum_{i=1}^{L}[(S^x_i)^2-(S^y_i)^2] \right),
    \label{eq:Haldane_anisotropic_hamiltonian}
\end{equation}
where $\bs{S}_i = (S^x_i, S^y_i, S^z_i)$ are spin-1 operators acting on 
site $i$. The model captures key features of quantum magnetism through 
the interplay of exchange interactions and anisotropies 
\cite{ren2018quantum}. The first sum describes an isotropic 
nearest-neighbor antiferromagnetic interaction with the exchange coupling $J>0$, which also sets the energy scale of the system, while the 
single-site terms are the axial and rhombic anisotropies ruled by the 
parameters $D$ and $E$~\cite{PhysRevB.96.060404}.
Both anisotropies play an essential role in defining the 
ground-state properties and the corresponding phase diagram of the 
model. The phase diagram of the model \eqref{eq:Haldane_anisotropic_hamiltonian} has been obtained through MPS analysis in Ref.~\cite{PhysRevB.96.060404} and is plotted in Fig.~\ref{fig:Cluster_Ising_model}(b). 
The model supports a topological phase, namely the central Haldane phase in the phase diagram, which is identified by a string order parameter \cite{den1989preroughening,ren2018quantum}. Moreover, it exhibits three antiferromagnetic 
(N\'eel) phases along the $x,y,$ and $z$ axes, and three trivial 
paramagnetic phases for large anisotropies. Our goal is to apply our SVM-based method to reproduce the phase diagram of the system. In Fig.~\ref{fig:Cluster_Ising_model}(b), we indicate the training points by  colored plus signs. The ground states are obtained 
by MPS using $\chi=50$.

\begin{figure*}[t]
	\centering
	\includegraphics[width=1\textwidth]{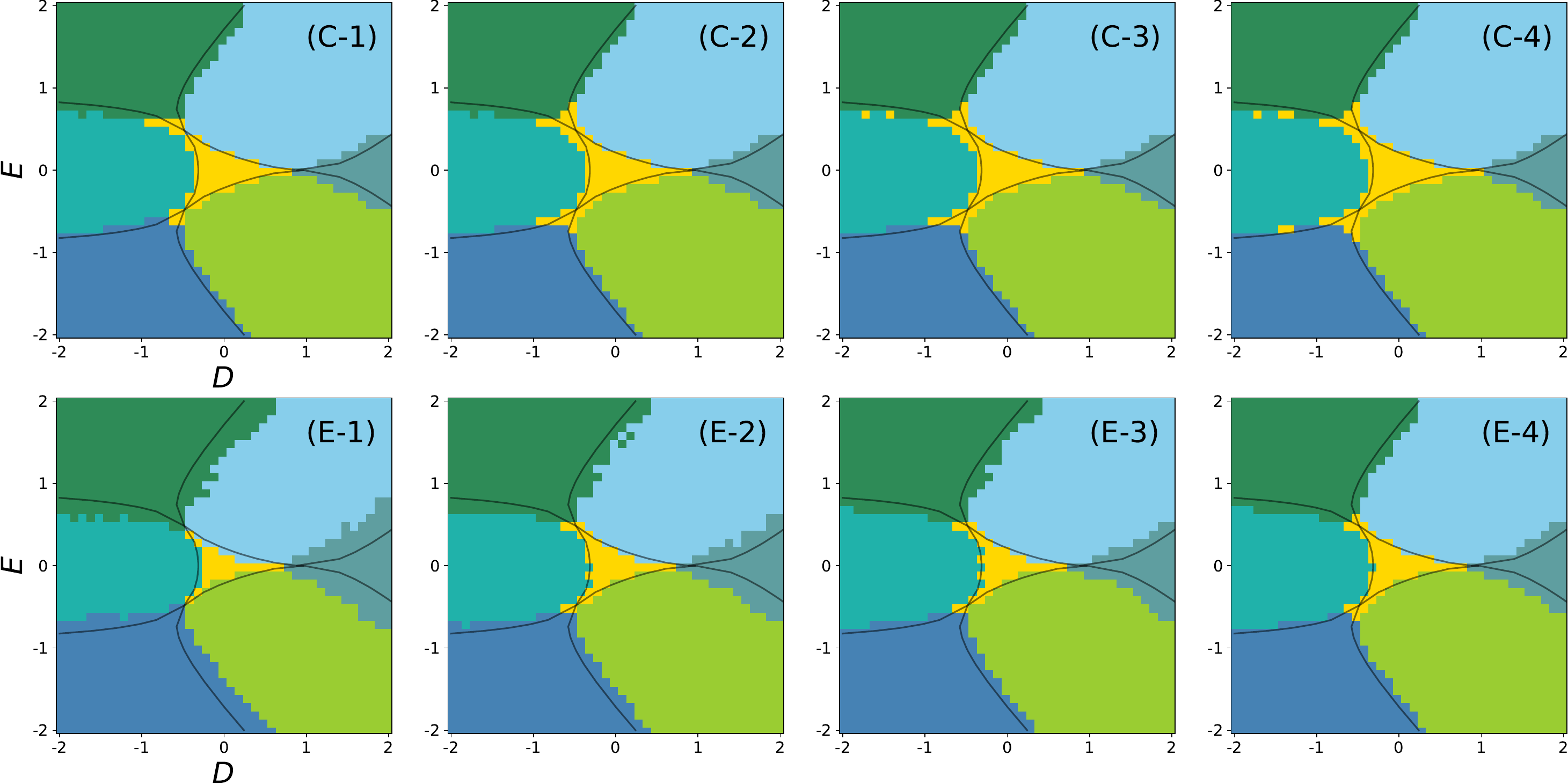}
  \caption{
Phase diagram reconstruction of the anisotropic Haldane model based on reduced density matrices for blocks of increasing number of sites. In the upper panels the block is at the center of the chain and (C-$k$) refers to keeping $k$ sites. In the lower panels the block is at the edge of the chain and (E-$k$) refers to keeping $k$ sites. A chain of 51 sites is used both for training and testing. 
%    Reconstruction of the phase diagram of the anisotropic Haldane model shown in Fig.~\ref{fig:Cluster_Ising_model}(b). Effect of keeping an increasing number of sites while tracing out the rest of the system. In the upper panel we keep the sites from the middle of the chain, and (C-$k$) refers to keeping $k$ sites. In the lower panel we keep the sites from an edge of the chain and (E-$k$) refers to keeping $k$ sites. For both training and testing  the full chain consists of 51 sites.
  }
  \label{fig:Haldane - 51 cite}
\end{figure*}
\begin{figure*}[th!]
	\centering
	\includegraphics[width=1\textwidth]{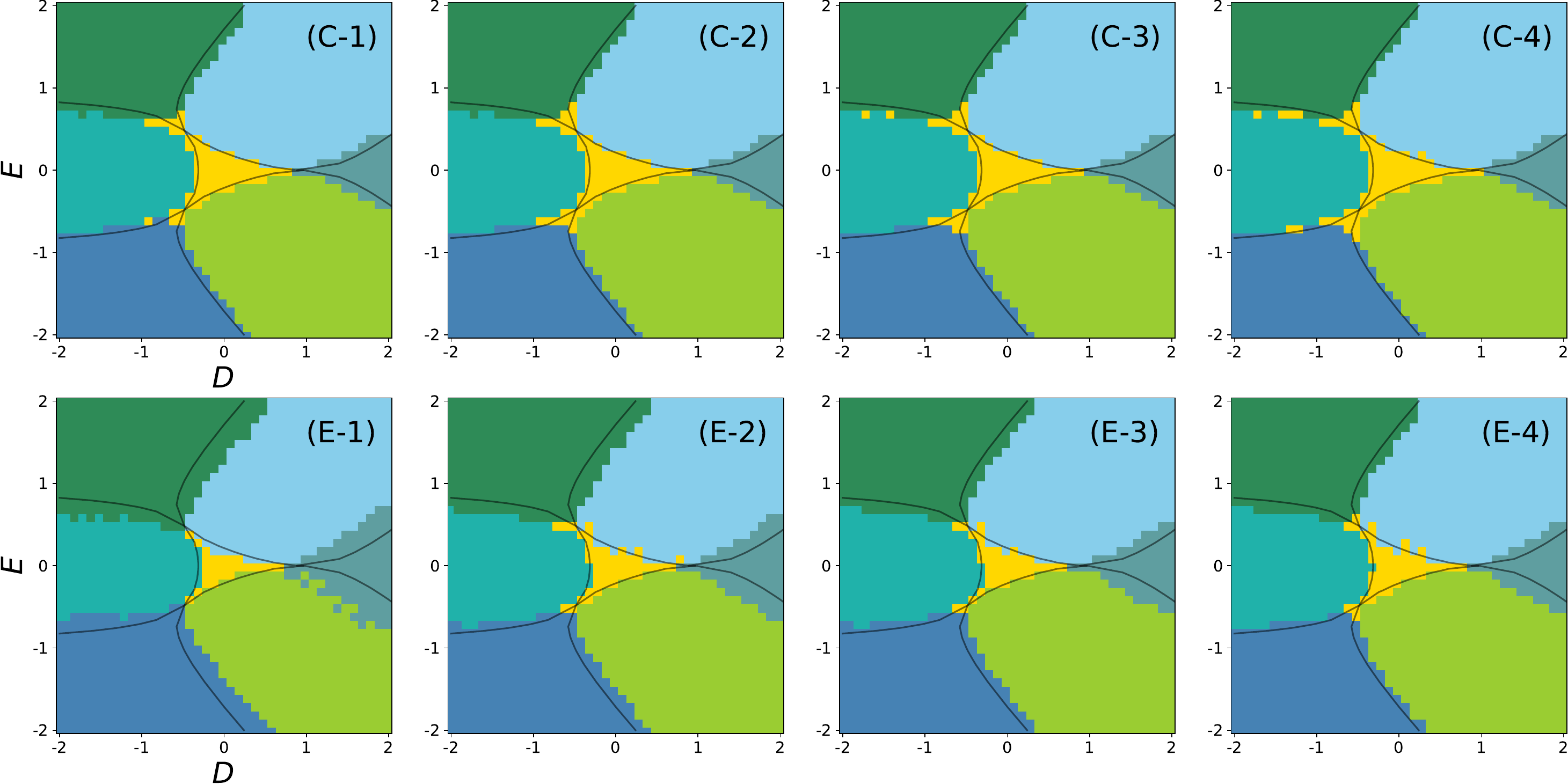}
  \caption{
    %Reconstruction of the phase diagram of the anisotropic Haldane model, as in Fig.~\ref{fig:Haldane - 51 cite}, but here training is performed with a chain of 31 sites, while for test we considered a larger chain of 51 sites.
Reconstruction of the phase diagram of the anisotropic Haldane model as in Fig.~\ref{fig:Haldane - 51 cite}; while a chain of 51 sites is still used for testing, training is performed with a shorter chain of 31 sites.
  }
  \label{fig:Haldane - 31 cite}
\end{figure*}

Figs.~\ref{fig:Haldane - 51 cite} and \ref{fig:Haldane - 31 cite} refer 
to the same analysis presented in Figs.~\ref{fig:Cluster Ising - 51} and 
\ref{fig:Cluster Ising - 31}, respectively, but now for the 
spin-1 chain Eq.~\eqref{eq:Haldane_anisotropic_hamiltonian}.
Panels (C-1) to (C-4) in Fig.~\ref{fig:Haldane - 51 cite} show that the phase diagram of the Haldane spin-$1$
chain can be reproduced with high accuracy by retaining only a small subsystem at the center of the chain -- in fact, even by considering just a single spin. The paramagnetic
phases are consistently identified whatever the size of the sections, 
whereas the recognition of N\'eel phases progressively improves as the 
sections get longer.
In panels (E-1) to (E-4) in Fig.~\ref{fig:Haldane - 51 cite}, we draw the phase diagram for the case that we keep the spins at the dge of che chain. As before, the results becomes worse when we keep the spins at the edge of the chain. In fact, a satisfactory recognition of the Haldane and N\'eel  phases is only 
obtained for the longest boundary sections, although the paramagnetic 
phases are more precisely classified across all cases, as expected.

Similar to the approach used for the generalized cluster-Ising spin-$1/2$ chain, one can train the model on smaller chains, let say $L=31$, and then deploy it for phase classification of longer ones, let say $L=51$. The results are shown in Fig.~\ref{fig:Haldane - 31 cite}. A comparison of Fig.~\ref{fig:Haldane - 51 cite} with Fig.~\ref{fig:Haldane - 31 cite} clearly shows comparable performance, indicating the generalizability of the model.

%\mehran{for the accuracy you can explain that by starting from edge the Haldane and neel phases are not distigushible that much then accuracy is low but when we keep from the middle from haldane phase we have a singlet spin which is truly distiguishable from the neel states. this is the reason why accuracy plot is like this}

%Figure \ref{fig: acc_vs_site_haldane}~(b) shows the accuracy of the topological phase. A similar trend is observed here. Keeping sites from the middle of the chain provides the best performance, reaching perfect accuracy even when only a single site is considered, and remaining at 100\% as more sites are added. In contrast, for the edge-site case, accuracy improves progressively with the number of kept sites. Both the 31-site and 51-site systems exhibit comparable behavior, with convergence achieved after approximately six sites are kept. \mehran{why accuracy reduced at site 3 }

\subsection{Overall performance}

\begin{figure}[t]
  \begin{center}
    \includegraphics[width=0.4\textwidth]{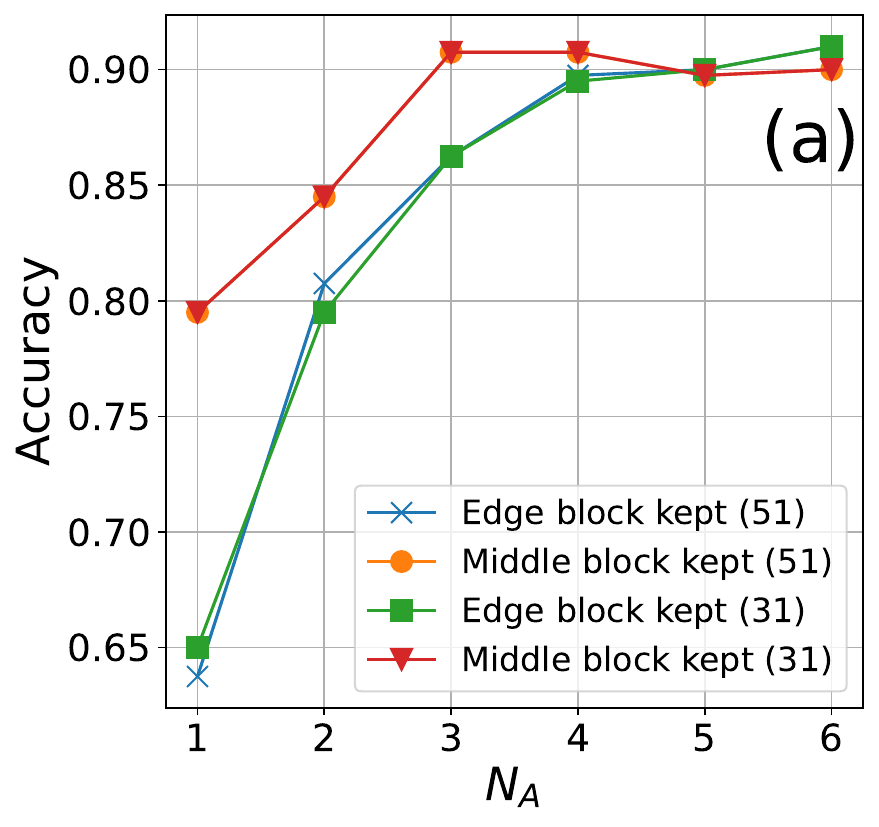}
    \includegraphics[width=0.4\textwidth]{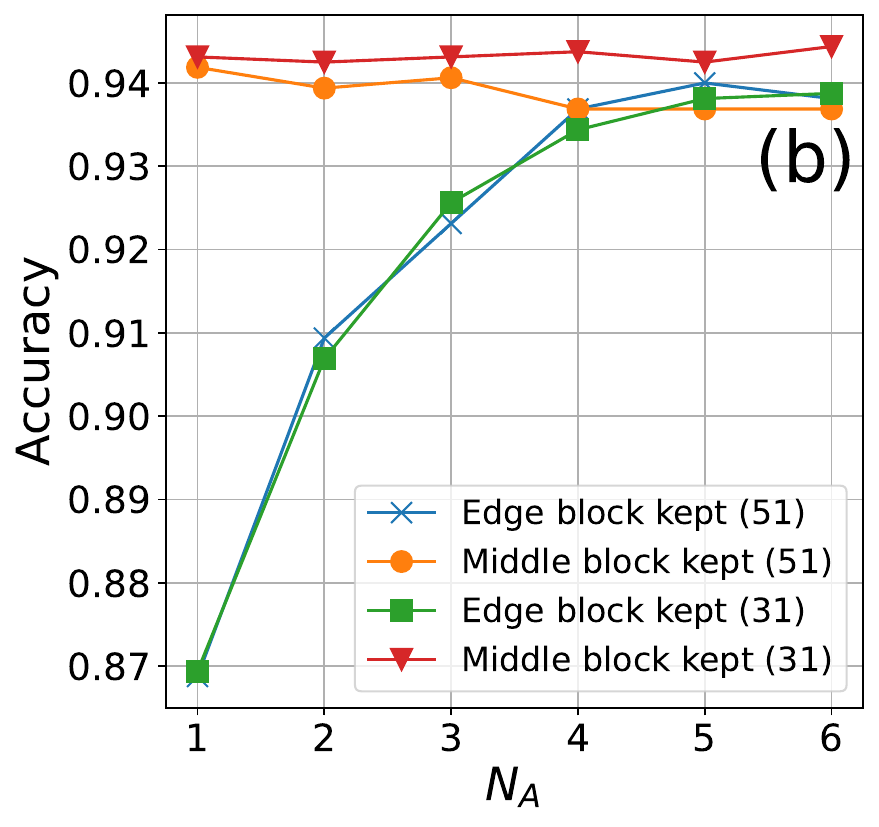}
    \caption{
      Accuracy in phase classification, namely fraction of correctly classified test points 
      versus number of kept sites, for (a) Cluster-Ising model and (b) anisotropic Haldane model. 
      Notice that in panel (a) the red and orange curves basically overlap. 
    }
    \label{fig: acc_vs_site}
  \end{center}
\end{figure}

The overall performance of our protocol, for both models,  is evaluated by computing its accuracy on the test points, which are taken as a regular grid spanning the entire phase diagram. The accuracy is defined as the ratio between the number of points properly assigned to the respective phase and the 
total number of tested points. In Figs.~\ref{fig: acc_vs_site}, we plot accuracy as a function of the number of qubits in block 
$A$, namely $N_A$, for chains of lengths $L=51$ and $L=31$, with the block placed either at the edge or at the center of the chain. The figures 
show that when the block $A$ is at the edge of the chain the accuracy monotonically improves by increasing the size of block $A$. However,  when the block $A$ is at the center of the chain the accuracy saturates as $N_A$ increases. In addition, when $N_A$ is very small the central block results in a better accuracy than the one at the edge but as $N_A$ increases the accuracies become comparable.  
Moreover, the analysis shows that systems of length 31 and 
51 produce almost identical accuracies. This indicates that the reduced density matrix of the subsystem has already converged and does not depend on the total system size.

About the effectiveness of our approach, we  first observe that a central section $A$ provides 
better results than an edge one. Then we see that sections of 
one single site allow recognition of paramagnetic phases, 
consistently with their being completely uncorrelated; on other other hand, for the cluster-Ising model 
such single-site sections are not sufficient to classify correlated ground states, for which at
least two contiguous spins are required to identify the 
antiferromagnetic phase, featuring an alternating 
spin alignment and two-site correlations.
Notably, the \ac{SPT} phase remains distinguishable even when A consists of a single spin. In fact, while a single-site reduced density matrix cannot directly capture non-local topological correlations, different physical phases occupy distinct regions within the space of reduced density matrices~\cite{zauner2016symmetry,henley2014density}. By training our supervised machine learning model across the phase diagram, the model effectively learns the decision boundaries defined by the alternative phases, allowing it to isolate the \ac{SPT} phase at the single-site level through contrast and exclusion.

\section{Conclusions \label{s.conclusion}}

In conclusion, we have demonstrated a quantum kernel framework capable of identifying rich phase diagrams of quantum many-body systems using information extracted solely from small  subsystems, even when they host topological phases. Our protocol utilizes a classical \ac{SVM} algorithm fed with a quantum kernel constructed from reduced density matrices, enabling the model to learn a specialized mapping that translates accessible local features to reconstruct global knowledge of the underlying phases. By benchmarking this protocol on the generalized cluster-Ising spin-1/2 chain and the anisotropic spin-1 Haldane chain, we verified that the reduced density matrices of as few as one to four sites retain sufficient features to identify all components of the phase diagram, including topological phases. Furthermore, the framework exhibits size-generalizability, reliably making predictions on longer chains even when trained on moderate-sized systems.

A central outcome of our work is the demonstration that the \ac{SVM} model can recognize topological phases without direct access to global string order parameters or full-system observables. This capability is fundamentally rooted in two main aspects: i) the supervised nature of our framework, which leverages our prior theoretical knowledge of the phase diagram during the training phase; and ii) the rich informational content extracted from the overlap of reduced density matrices that defines the quantum kernel. By shifting the burden from the demanding measurement of global observables to the estimation of a local kernel matrix, our methodology offers a practical, low-overhead route for characterizing the phase diagrams of strongly correlated many-body systems.

\section*{Acknowledgment}
We thank Yu-Chin Tzeng for generously providing the data used to reconstruct Fig. \ref{fig:Cluster_Ising_model}(b) and for valuable discussions that significantly contributed to this work.
M.K.,~P.V.,~A.C.~and L.B.~acknowledge financial support from: PNRR Ministero Universit`a e Ricerca Project No. PE0000023-NQSTI funded by European Union-Next-Generation EU.
A.C.~and L.B.~are supported by the QUART\&T project funded by the
Italian Institute of Nuclear Physics (INFN) within the Technological and Interdisciplinary Research Commission (CSN5).
L.B.~also acknowledges financial support from: Prin 2022 - DD N. 104 del 2/2/2022, entitled ``understanding the LEarning process of QUantum Neural networks (LeQun)'', proposal code 2022WHZ5XH, CUP B53D23009530006; the European Union's Horizon Europe research and innovation program under EPIQUE Project GA No. 101135288. A.B.~acknowledges support from the National Natural Science Foundation of China (grants No. W2541020, No. 12274059, No. 12574528 and No. 1251101297).

\bibliography{biblio}

\end{document}